\documentclass[%
 amsmath,amssymb,
preprint,%
 graphicx,
]{revtex4}

\usepackage{graphicx}
\usepackage{dcolumn}
\usepackage{bm}

\usepackage[utf8]{inputenc}
\usepackage[T1]{fontenc}
\usepackage{mathptmx}
\usepackage{hyperref}
\usepackage{color}

\begin{document}



\title{Ultra-high-speed color imaging with single-pixel detectors under low light level} 



\author{Weigang Zhao}
 \affiliation{Electronic Materials Research Laboratory, Key Laboratory of the Ministry of Education \& International Center for Dielectric Research, Xi’an Jiaotong University, 28 West Xianning Road, Xi’an, China}
\author{Hui Chen}
 \affiliation{Electronic Materials Research Laboratory, Key Laboratory of the Ministry of Education \& International Center for Dielectric Research, Xi’an Jiaotong University, 28 West Xianning Road, Xi’an, China}
\author{Yuan Yuan}
 \email{yuanyuan0@xjtu.edu.cn}
 \affiliation{Electronic Materials Research Laboratory, Key Laboratory of the Ministry of Education \& International Center for Dielectric Research, Xi’an Jiaotong University, 28 West Xianning Road, Xi’an, China}
\author{Huaibin Zheng}
 \affiliation{Electronic Materials Research Laboratory, Key Laboratory of the Ministry of Education \& International Center for Dielectric Research, Xi’an Jiaotong University, 28 West Xianning Road, Xi’an, China}
\author{Jianbin Liu}
 \affiliation{Electronic Materials Research Laboratory, Key Laboratory of the Ministry of Education \& International Center for Dielectric Research, Xi’an Jiaotong University, 28 West Xianning Road, Xi’an, China}
\author{Jianbin Liu}
 \affiliation{Electronic Materials Research Laboratory, Key Laboratory of the Ministry of Education \& International Center for Dielectric Research, Xi’an Jiaotong University, 28 West Xianning Road, Xi’an, China} 
\author{Yu Zhou}
 \affiliation{MOE Key Laboratory for Nonequilibrium Synthesis and Modulation of Condensed Matter,
Department of Applied Physics, Xi’an Jiaotong University, 28 West Xianning Road, Xi’an, China}
\author{Zhuo Xu}
 \affiliation{Electronic Materials Research Laboratory, Key Laboratory of the Ministry of Education \& International Center for Dielectric Research, Xi’an Jiaotong University, 28 West Xianning Road, Xi’an, China}


\date{\today}

\begin{abstract}
Single-pixel imaging is suitable for low light level scenarios because a bucket detector is employed to maximally collect the light from an object. However, one of the challenges is its slow imaging speed, mainly due to the slow light modulation technique. We here demonstrate 1.4MHz video imaging based on computational ghost imaging with a RGB LED array having a full-range frame rate up to 100MHz. With this method, the motion of a high speed propeller is observed. Moreover, by exploiting single-photon detectors to increase the detection efficiency, this method is developed for ultra-high-speed imaging under low light level.
\end{abstract}

\pacs{}

\maketitle 

\section{Introduction}

Ultra-high speed imaging under low light level is an indispensable diagnostic tool for many disciplines and applications, such as detecting bio-dynamics in living cells or observing microfluidics \cite{Goda2009Serial,NakagawaSequentially,Nakagawa2015Sequentially}. It requires an imager having capabilities of fast storing and processing large data flow. Simultaneously, that imager should be sensitive enough to detect a dynamic event within an ultra-short time period. Nevertheless, to meet both these requirements is challenging. For repetitive dynamic events, the second requirement can be worked round by repeatedly measuring the same event within an ultra-short time period until the accumulated light energy is adequate to reconstruct a frame. Various high-speed imaging techniques for repetitive dynamic events have been developed over recent years, such as pump–probe technique, the framing streak camera and tracing the flight of laser \cite{Gariepy2015Single,Hockett2011Time,Wong2012Electronic,Acremann2000Imaging,Feurer2003Spatiotemporal,Kodama2001Fast,Velten2012Recovering,Hajdu2000Analyzing,Zewail2010Four,Barty2010Ultrafast}. However, these techniques are unable to capture unique or random events (such as molecular motion in biological samples), limiting their uses in practical applications. 

To image a non-repetitive ultra-fast event, a method must have capability of continually recording images at a high frame rate, and as well as high sensitivity to capture enough light within a short time. Conventional spatial resolving detectors such as  charge-coupled devices (CCDs) and complementary metal–oxide–semiconductor (CMOS) are not suitable for ultra-high speed imaging under low light level \cite{Etoh2017TheCCD}. It is due to the compromise between the response speed of pixelated sensors and their sensitivity. Although a CCD imager with 1MHz frame rate has been developed, it requires a high power illumination to acquire enough light for each pixel within the short time of a frame\cite{Etoh2007Evolution}. This bright illumination may damage samples in some scenarios such as biological imaging.

Single-pixel imaging (SPI) method is an alternative way to realize ultra-high speed imaging\cite{Bian2016MultispectralSPI,Higham2018DeepSPI,SPIviaCS,Radwell2014SPI,Goda2009Serial,Nakagawa2015Sequentially,NakagawaSequentially}. It uses a temporal measurement approach to replace the spatial resolving detection scheme, and then employs an ultra-fast and sensitive single detector to detect the light variance that was modulated by an object. One of the high-speed SPI methods is using photonic time stretch (PTS) technique, also named dispersive Fourier transform and frequency-to-time mapping. PTS technique stretches the broadband optical spectrum of an ultra-short laser pulse, and maps it into 2D spatial waveform that is projected onto the object. The field reflected from the object is recombined as (transformed into) a temporal waveform that is able to be measured with a high-speed single-pixel detector\cite{Nakagawa2015Sequentially,NakagawaSequentially}. This technique has several drawbacks: (1) in each frame, the 2D spatial reflection light intensities are encoded into a temporal waveform, and the single-pixel detector is used to decode the spatial signals, which does not essentially increase detection sensitivity; (2) it is of high cost due to the expensive ultra-short pulsed laser and the optical amplifier; (3) it works only at certain optical spectra, which limits its applications such as in color imaging, or imaging for some materials that has bad response at the probing spectrum.

Ghost imaging and single-pixel camera are another two types of single-pixel imaging schemes, which use bucket detectors to collect the maximum level of the light from an object\cite{Pittman1995OpticalGIfirst,Zhang2005Correlatedthermallight,Shapiro2008ComputationalGI,Bromberg2008GhostSPD}. This detection scheme architecturally enables much higher sensitivity than spatial resolving detecting schemes (such as CCD and PTS). Ghost imaging has been applied to X-ray and is capable of imaging under ultra-low radiation\cite{Pelliccia2016firstXrayGI,Yu2016FourierXray,AiTabletoplowdoseXray,Schori2017Xray}. On the other hand, this scheme requires thousands of measurements to recover a frame, resulting in dramatically demanding much higher detection speed than those mentioned-above techniques, raising the technical difficulty. 

Previously, the usage of spatial light modulators (SLMs) or digital micromirror devices (DMDs) limit the imaging speed of computational ghost imaging (CGI)\cite{Shapiro2008ComputationalGI,Bromberg2008GhostSPD,Lu2011GhostDMD}. Our group proposed the idea of using high-speed modulation LED array as a light source to boost the imaging speed in a patent in 2017\cite{IEEE22}. A similar research is independently announced in 2018\cite{Xu20181000}. We here develop a method to overcome the compromise between the sensitivity and high frame rate, and enable ultra-high-speed imaging under low light level. As a proof of this proposal, we demonstrate the following experiments: (1) imaging at 1.4MHz frame rate for a fast rotating propeller sweeping a letter; (2) high-speed imaging for a colored object; (3) high-speed imaging under low light level with single-photon detectors to boost the detection sensitivity, where the illumination on the object is less than $3\times10^9$ photons per second per $mm^2$.

\section{Experiment}

\begin{figure}
    \centering[htbp]
    \includegraphics[width=150mm]{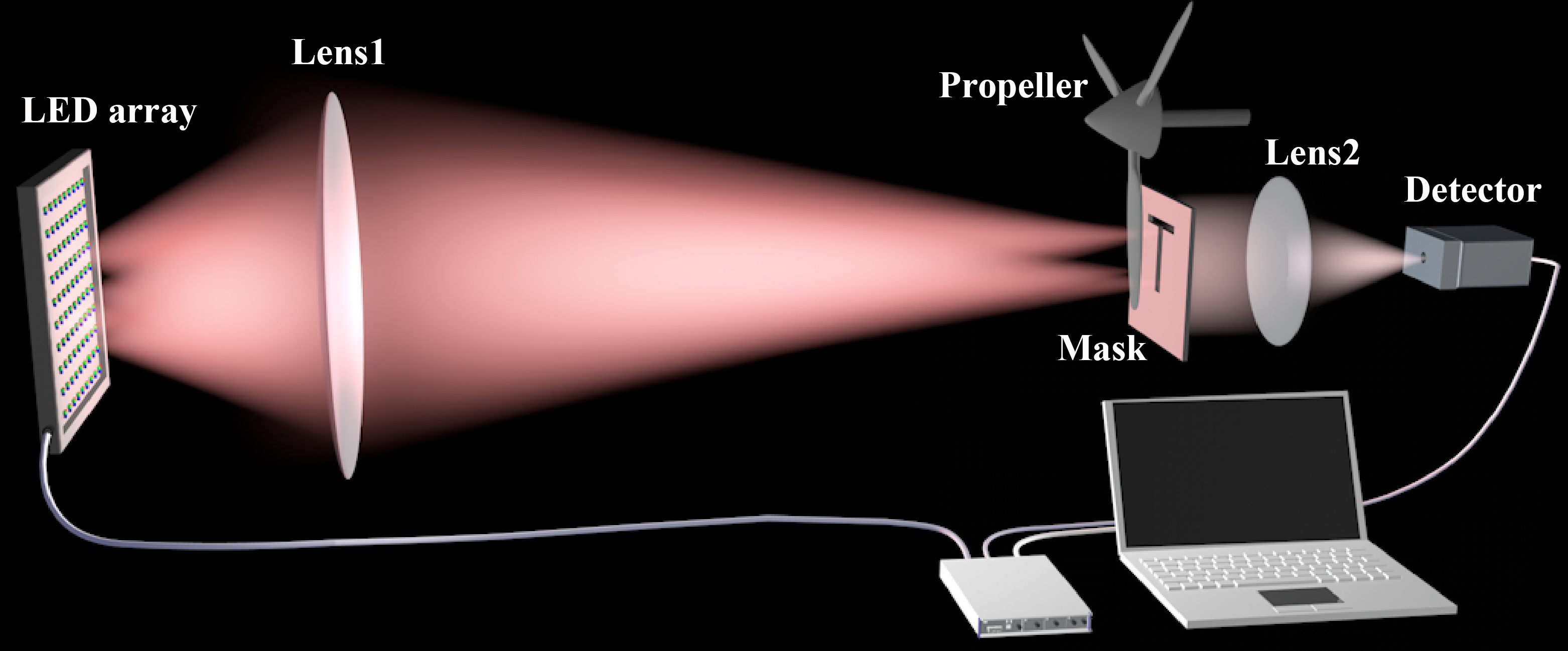}
    \caption{The schematic of the experimental setup for a moving detection.  The object consists of a static mask and a rotating propeller at a speed of 40k RPM. Lens1 is used to project the illumination patterns onto the object. The transmitted light from the object is collected by a PMT detector via a focusing lens (Lens2).}
    \label{fig:my_label1}
\end{figure}

The experimental setup is shown in Figure \ref{fig:my_label1}. The light source is a self-made $10\times10$ LED array. Each element consists of a red, a green and a blue LED light bulbs. A FPGA circuit was designed to simultaneously control all the LED bulbs on or off with a full frame rate up to 100MHz. A sequence of arbitrary patterns can be loaded into the memory of the circuit from a computer via a USB port. Afterwards, the stored patterns are emitted with a time interval of 10 $ns$ (100MHz), and projected onto an object via a lens (L1).

In the first experiment, a full set of $8 \times 8$ Hadamard patterns is pre-stored into the circuit. Right ahead of the Hadamard patterns, we place a mark that consists of three sets of full white and full black patterns, which helps to precisely determine the staring time of sending the patterns. All the patterns ($64+6=70$ in total, 6 represents the flag patterns) are successively displayed. After the display cycle ends, another cycle will keep repeating until the end of the experiment. The light distribution on the object plane can be formulated as
\begin{equation}
    M(x,y;t)=\sum_{j}P_j(x,y)\cdot W_{\tau}(t-T_{m,j})  
    \label{equation1}
\end{equation}
Here $P_j(x,y)$ is the $j$-th projected pattern. $W_{\tau}(t-T_{m,j})$ is a window function: it is one over an interval of $[T_{m,j},T_{m,j}+\tau]$, but zero outside this region. $\tau$ is the lasting time of a pattern. $T_{m,j}=m\cdot T_{c}+j\times T_{j}$, is the starting time of the $j$-th pattern in the $m$-th cycle, where $T_{c}$ is the total time of a cycle and $T_j$ is the time interval between two adjacent patterns.

The object is a hollow letter "T", right in front of which is propeller rotating at 40000 RPM (rev. per min.). The length of the propeller is about 5 $cm$. Thus, the linear speed at the edge is about 200 $ m/s$. 
The transmitted light through the object is collected by a bucket detector (a PMT with a 250MHz bandwidth). The bucket detector senses the intensity variance that represents how the projected patterns were modulated by the object
\begin{equation}
   B(t)= \iint_{s} M(x,y;t)\cdot O(x,y;t)dx dy   
   \label{equation2}
\end{equation}
Here, $O(x,y;t)$ is the intensity attenuation function of the object. The integral indicates all the transmitted light from the object collected by the bucket detector. The second-order correlation is calculated as
\begin{equation}
    G(x,y;t=m\cdot T_c)=\int_{t}^{t+T_\tau} M(x,y;t')\cdot B(t') dt' 
    \label{equation3} 
\end{equation}
Which exhibits the image of a dynamic object at time $t=m\cdot T_c$. $t=m\cdot T_c$ represents that we reconstruct an image frame in every cycle. With the help of mark, we are able to locate the time of the first pattern from the bucket signal which is used to synchronize the times of the bucket signal and the projected patterns. 

\begin{figure}[htbp]
    \centering
    \includegraphics[width=150mm]{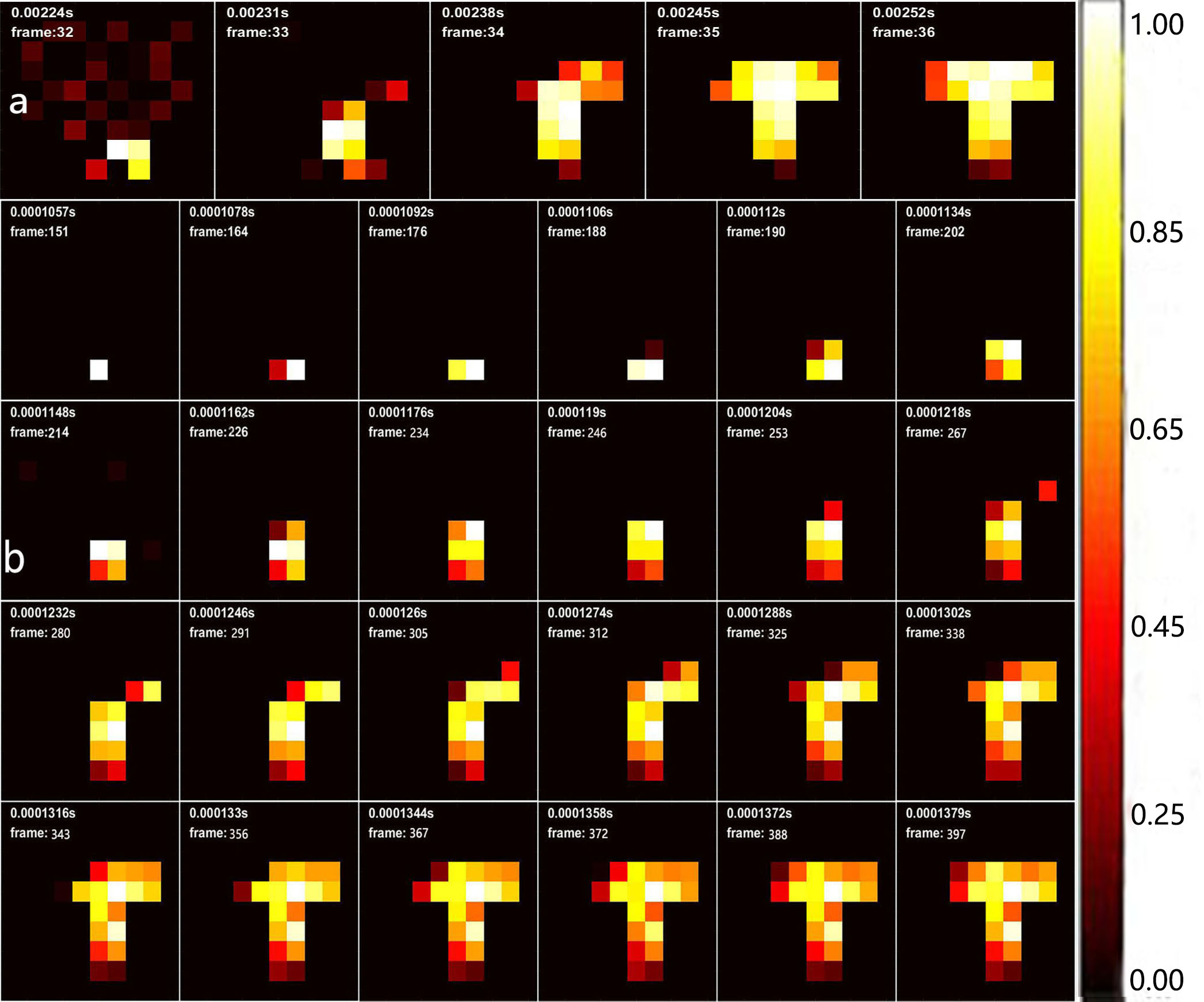}
    \caption{The video frame of the imaging results. (a) Imaging video with a 1MHz speckle modulation rate; (b) Video with 100MHz modulation rate. Note that, the images are enhanced with a background removal process.}
    \label{fig:my_label2}
\end{figure}

The motion of the propeller swapping "T" is captured with frame rate of 0.014MHz and 1.4MHz (1MHz and 100MHz modulation rates), as shown in Figure \ref{fig:my_label2}a and \ref{fig:my_label2}b, respectively. Please watch \href{https://weibo.com/tv/v/Hq4J4emSh?fid=1034:4362146935426315}{\color{blue} ac1MHz.avi} and \href{https://weibo.com/tv/v/Hq4Dn4nnR?fid=1034:4362143177350319}{\color{blue}ac100MHz.avi} for the videos. As shown in Figure \ref{fig:my_label2}a frame by frame, the imaging speed of 0.014MHz is not fast enough the capture all the motion of the propeller. Therefore, the 1.4MHz frame rate is necessary for such a fast motion.

\begin{figure}[htbp]
    \centering
    \includegraphics[width=150mm]{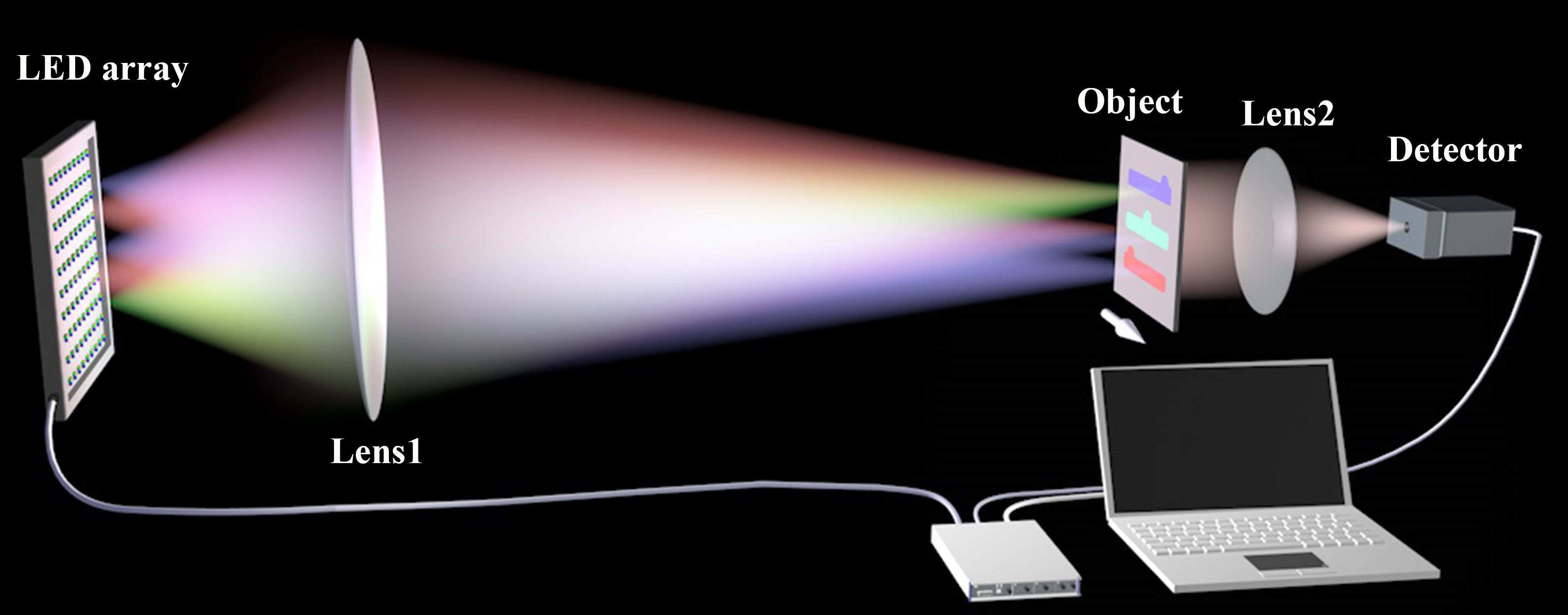}
    \caption{The schematic of the experimental setup for a colorful object.  The object consists of three strip-like objects in red, green, and blue, respectively. Lens1 is used to project the illumination patterns onto the object. The transmitted light from the object is collected into a PMT detector via a focusing lens (Lens2). }
    \label{fig:my_label3}
\end{figure}

\begin{figure}[htbp]
    \centering
    \includegraphics[width=150mm]{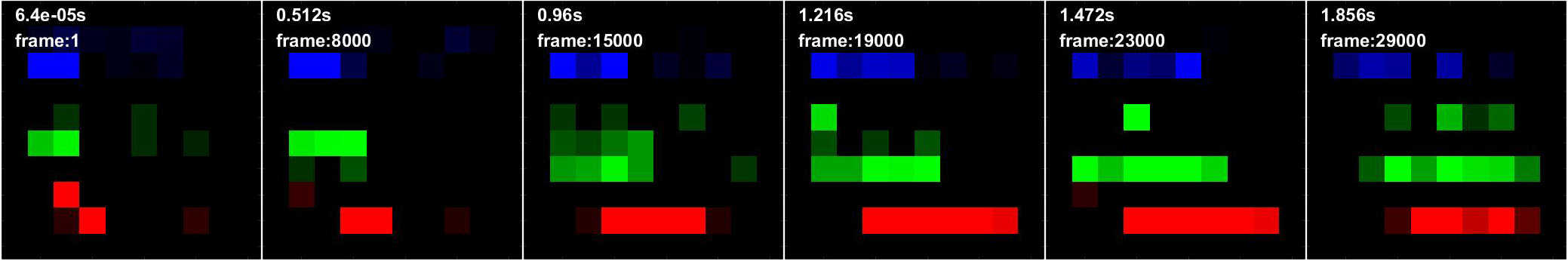}
    \caption{The video screenshots from color imaging process. When the imager is capturing images, the object moves from left to right.}
    \label{fig:my_label4}
\end{figure}

To perform color imaging, each Hadamard pattern is repeated triple times in red, green and blue, respectively. Therefore, the total sub-patterns in a cycle is $6+64\times3=198$.  The object consists of three plates in the three colors, as shown in Figure \ref{fig:my_label3}. The object is placed on a moving stage moving from left to right. Note that the blue illumination of our source is weaker than the green and the red ones, and the detector is also less sensitive for the blue light. To achieve a better imaging quality,the LED array was set to work at a frame rate of 1MHz. The imaging frame rate becomes 5kHz. The selected video frames are shown Figure \ref{fig:my_label4}. Please see \href{https://weibo.com/tv/v/Hq4HJpYS8?fid=1034:4362144209161039}{\color{blue}color.avi} for the whole video.

\begin{figure}[htbp]
    \centering
    \includegraphics[width=150mm]{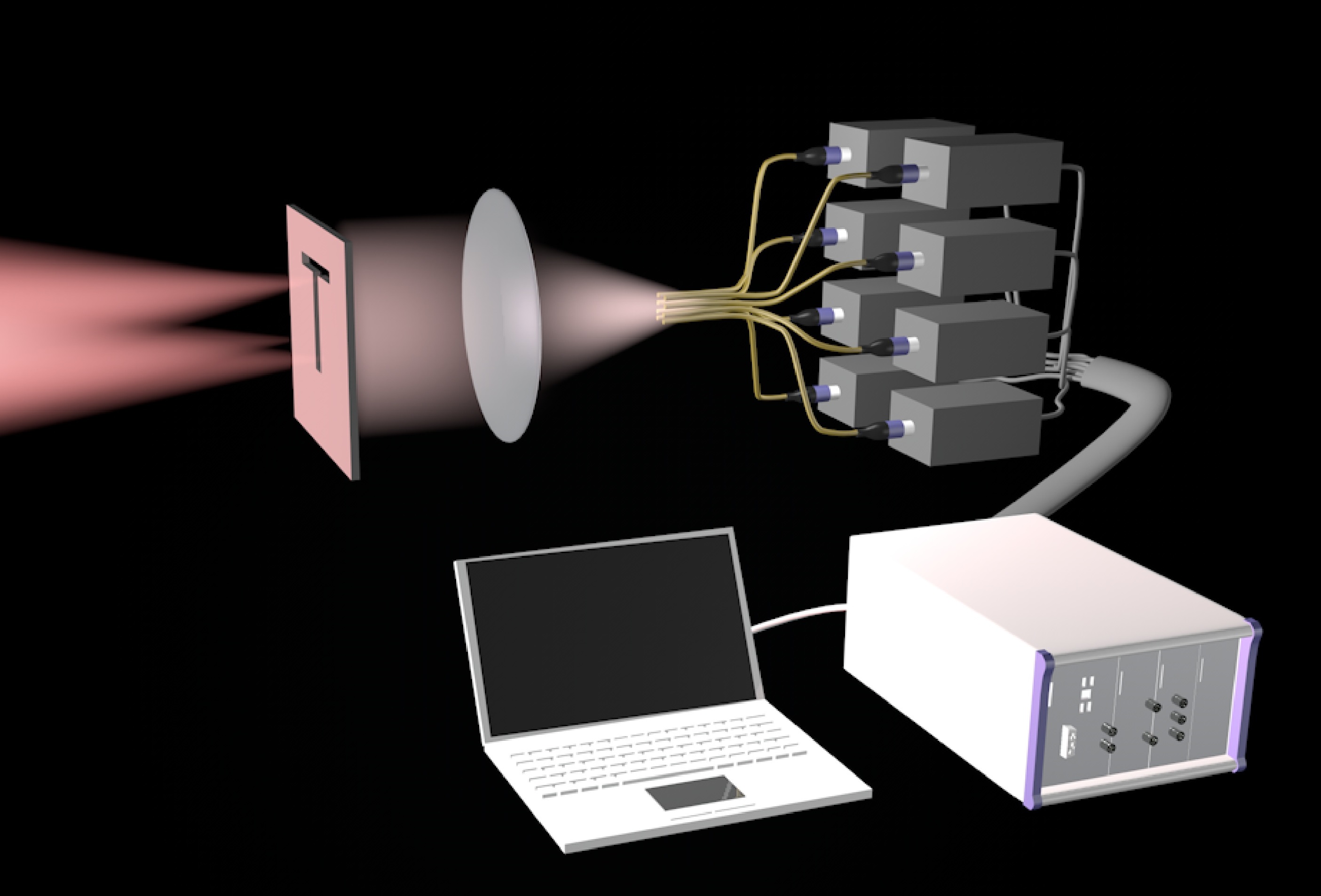}
    \caption{The low-light-level imaging experimental scheme. The transmitted light from the object is collected by eight multi-mode fibers via a lens. Each fiber is connected with a single-photon detector to count photon number reaching the detection plane. }
    \label{fig:my_label5}
\end{figure}

Under low light level, the light intensity within a short integral time window is weak. To maximally collect the weak light from the object, we employed eight single-photon detectors. As shown in Figure \ref{fig:my_label5}, eight multi-mode fibers were close-packed at one end. At another end, each fiber was connected to a single-photon detector. The signals output from the detectors were fed into a time-correlated single photon counting (TCSPC) device, which recorded the arrival times of photons detected by the detectors. Therefore, the photon number arriving in each integral time window can be counted by analyzing the photons’ arrival time. This setup dramatically increases the detection sensitivity, allowing us implement the ultra-high speed imaging under a certain low light level.  

\begin{figure}[htbp]
    \centering
    \includegraphics[width=80mm]{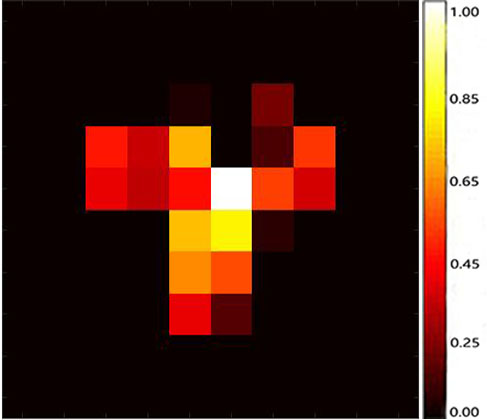}
    \caption{The image of a letter "T" recovered within 700 $\mu s$. Eight single-photon detectors are used to collect the low photon flow under 1MHz modulation rate. }
    \label{fig:my_label6}
\end{figure}

In the experiment as shown in Figure \ref{fig:my_label5}, we attenuate the illumination light to a level that the light intensity falling onto the detection plane is $3\times10^9 photons \cdot s^{-1} \cdot mm^{-2}$. Note that, under such a low photon flux, the PMT detector used in our previous experiments failed to yield usable signals. Since the single-photon detectors have a dead time of 80 $n s$, the imager runs at 1MHz modulation rate. On the  other hand, the average photon number in 1 $\mu s$ is less than 50. To achieve a better statistical result, we thus use 10 cycles to recover a frame, i.e., the time of an imaging frame is 700 $\mu s$. A recovered imaging frame of a letter “T” is shown in Figure \ref{fig:my_label6}.

The imaging speed is limited by the average photon number per modulation frame. Due to the long dead time, eight single-photon detectors are not enough to prove  sufficient dynamic range of detection, which therefore caused the lower imaging speed of above experiment. However, this limitation can be easily overcome by modern detection technology. Commercial single-photon detector array($32 \times 32$\cite{Richardson2009SPADarray} and even $256\times 256$ \cite{Luca2015ASPAD256}) has been announced and used in many fields, which can provide a dynamic range more than 40000, with which the imaging speed can be boosted to 1MHz and higher.

\begin{figure}[htbp]
    \centering
    \includegraphics[width=80mm]{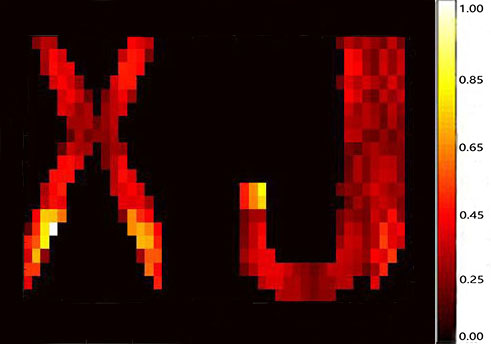}
    \caption{Image of the object of letter 'XJ', which is reconstructed in $ 126 \mu s$ with 18 frames under 10 MHz modulation rate.}
    \label{fig:my_label7}
\end{figure}

Currently, the array of our illumination device is only $10 \times 10$, limiting its applications. Nevertheless, the resolution can be extended without changing the hardware configuration in some situations. Usually, the motion speed of an real object is quite slow in comparison with the speed of the modulation rate. A high resolution image can be obtained by scanning an object with illumination patterns. In experiment, an object of “XJ” is mounted on a 2D motorized stage, then the object was scanned by the illuminating pattern horizontally and vertically recording the coordination at the same time. In each scan, 1/18 of the object is imaged within 7 $\mu s$ (under 10MHz modulation rate). After scanning, an image with high resolution of $24\times 48$ is retrieved, taking 126  $ \mu s$ in total. Figure \ref{fig:my_label7} shows the reconstruction image of the whole object.

\section{Discussion}

Although the modulation rate of electronic circuit can higher than 100MHz, the LED bulbs have the rising and falling times limits the highest modulation speed. If the electronic modulation rate is higher than the speed of LED bulbs, the variance of the emission light can not catch up with the electronic modulation. For example, if the modulation rate is 1GHz but the rising time of the light bulb is 10ns, when the circuit tries to turn a light bulb on in 1 ns and then off in the next 1 ns, the light bulb would not correctly response this modulation. In order to solve this shortage, we propose a method of replacing the RGB LED bulb with laser diode(LD), which can be modulated at pico-second time scale. At the same time, the state of art in semiconductor industry is easily to fabricate a light emission array with much higher resolution than $10\times10$ and a modulation rate faster than 1GHz.

In this article, we experimentally demonstrate an ultra-high-speed color imaging system, with which the high-speed motion of the objects is observed. In particular, with single-photon detectors, the imager’s sensitivity is enhanced, enabling imaging high-speed moving objects under low light level. Our work extends the applications of ghost imaging to high-speed diagnostics in low light scenarios, and will inspire the other applications. 

\section*{Funding Information}    
National Science Foundation (NSF) (11503020);



\bibliography{LED}

\end{document}